\documentclass[10pt]{amsart}
\newtheorem{thm}{Theorem}
\newtheorem{cor}[thm]{Corollary}
\newtheorem{lem}[thm]{Lemma}
\newtheorem{sat}[thm]{Proposition}
\theoremstyle{definition}

\usepackage{natbib}
\bibpunct{(}{)}{;}{a}{,}{,}

\usepackage{srcltx}
\usepackage{amssymb,color}

\definecolor{c10}{rgb}{0.1,0.5,0.4}
\definecolor{c15}{rgb}{0.6,0.1,0.4}
\definecolor{c20}{rgb}{0.,0.7,0.}
\definecolor{c30}{rgb}{0.,0.,1.}
\definecolor{c40}{rgb}{1,0.1,0.7}
\definecolor{c50}{rgb}{1,0,0}

\def\cEE#1{\textcolor{c10}{#1}}
\def\cEE#1{#1}

\def\cc#1{\textcolor{c30}{#1}}
\def\cc#1{#1}

\def\cb#1{\textcolor{c30}{#1}}
\def\cb#1{#1}
\def\cbb#1{\textcolor{c20}{#1}}
\def\cbb#1{#1}

\def\coB#1{\textcolor{c50}{#1}}
\def\coB#1{#1}
\def\cob#1{\textcolor{c50}{#1}}
\def\cob#1{#1}

\def\coC#1{\textcolor{c30}{#1}}
\def\coC#1{#1}

\def\coE#1{\textcolor{c20}{#1}}
\def\coE#1{#1}

\def\rL#1{\textcolor{c50}{#1}}
\def\rL#1{#1}

\def\coEH#1{#1}

\def\cJ#1{\textcolor{c50}{#1}}
\def\cJ#1{#1}

\newcommand{\nwc}{\newcommand}
\nwc{\COM}[1]{}

  \def\Rset{\mathbb{R}}

\newcommand{\netheo}[1]{{Theorem \ref{#1}}}

\newcommand{\kb}[1]{\boldsymbol{#1}}
\newcommand{\vk}[1]{\kb{#1}}

\def\FRE{\mbox{Fr\'{e}chet }}

\def\x{\vk{x}}

\def\X{\vk{X}}

\newcommand{\abs}[1]{\lvert #1 \rvert}

\newcommand{\E}[1]{\mbox{\rm$\vk{E}$}\{#1\}}
\newcommand{\eg}[1]{\mbox{\rm$\vk{E}$}\biggl\{#1 \biggr\}}

\newcommand{\pk}[1]{\mbox{\rm$\vk{P}$} \{#1\} }
\newcommand{\pb}[1]{\mbox{\rm$\vk{P}$}\Bigl \{#1 \Bigr \}}

\def\R{\Rset}

\newcommand{\inr}{\in \R}

\newcommand{\ldot}{,\ldots,}

\newcommand{\limit}[1]{\lim_{#1 \to   \infty}}

\newcommand{\todis}{\stackrel{d}{\to}}

\newcommand{\equaldis}{\stackrel{d}{=}}

\newcommand{\QED}{\hfill $\Box$}
\newcommand{\IF}{\infty}
\def\kal#1{{{ $\cal #1$}}}

\newcommand{\BQN}{\begin{eqnarray}}
\newcommand{\EQN}{\end{eqnarray}}
\newcommand{\BQNY}{\begin{eqnarray*}}
\newcommand{\EQNY}{\end{eqnarray*}}
\newcommand{\BS}{\begin{sat}}
\newcommand{\ES}{\end{sat}}
\newcommand{\BL}{\begin{lem}}
\newcommand{\EL}{\end{lem}}
\newcommand{\BT}{\begin{thm}}
\newcommand{\ET}{\end{thm}}
\newcommand{\BK}{\begin{cor}}
\newcommand{\EK}{\end{cor}}
\newcommand{\BD}{\begin{de}}
\newcommand{\ED}{\end{de}}
\newcommand{\BIT}{\begin{itemize}}
\newcommand{\EIT}{\end{itemize}}
\newcommand{\BDI}{\begin{description}}
\newcommand{\EDI}{\end{description}}

\newcommand{\BEX}{\begin{exxa}}
\newcommand{\EEX}{\end{exxa}}

\def\kal#1{\mathcal{#1}}

\def\X{\vk{X}}
\def\Y{\vk{Y}}

\begin{document}

\title[Random Shifting and Scaling]
{Random Shifting and Scaling of Insurance Risks}

\author{Enkelejd  Hashorva }
\address{Enkelejd Hashorva, Department of Actuarial Science\\Faculty of Business and Economics\\
University of Lausanne\\
B\^{a}timent Extranef, UNIL-Dorigny, 1015 Lausanne, Switzerland
}
\email{Enkelejd.Hashorva@unil.ch}

\author{Lanpeng Ji }
\address{Lanpeng Ji, Department of Actuarial Science\\Faculty of Business and Economics\\
University of Lausanne\\
B\^{a}timent Extranef, UNIL-Dorigny, 1015 Lausanne, Switzerland
}
\email{Lanpeng.Ji@unil.ch}

\subjclass[2000]{}


\begin{abstract}
Random shifting typically appears in credibility models whereas random scaling is often encountered in stochastic models for \cJ{ claim sizes} reflecting the time-value property of money. In this article we discuss some aspects of  random shifting  and  random scaling in insurance focusing in particular on credibility models, dependence structure of claim sizes in collective \cc{risk models}, and extreme value models for the joint dependence of large \cc{losses}.
We show that specifying certain actuarial models using random shifting or scaling has some advantages for both theoretical \cc{treatments} and practical applications.
\COM{A key finding of this contribution is that,
if we model claim sizes of an insurance portfolio by a  Dirichlet random sequence, then it turns out that this dependence model  can be casted in the framework of a random scale model of independent claims with known distribution functions. Further, with motivation from credibility theory, we introduce \cJ{a} new class of $L_P$ Dirichlet random vectors which is also related to \cJ{random scale models} and can be used for different modeling purposes.
}
\end{abstract}

\maketitle


\section{Introduction}
Random shifting and random scaling in \cJ{insurance applications are} natural phenomena \rL{for}  latent unknown risk factors, time-value of money, or the need of allowing financial risks to be dependent.
In this contribution, we are concerned with three principal stochastic models related to credibility theory, ruin theory, and extreme value modeling of  large losses.

In credibility theory  (e.g., \citet{Denetal2006}) often stochastic models are defined via a conditioning argument.
As an illustration, consider the classical Gaussian model assuming that the conditional random variable  $X \lvert \Theta=  \theta $ has the normal distribution $\mathcal{N}(\theta, \sigma^2)$. If further the random variable $\Theta$ has the normal distribution  $\mathcal{N}(\mu, \tau^2)$,
 we obtain the credibility premium formula for the Bayesian premium (calculated under the $L_2$ loss function)
 \BQN
\label{eqn:BI}
\E{\Theta \lvert X=x}  &=& x+\frac{\sigma^2}{\sigma^2+ \tau^2} \coEH{(\mu -x)}
\EQN
for any $x, \mu\in\R$ and $\sigma, \tau$ positive. 
 The relation explained by \eqref{eqn:BI} can  be directly derived by introducing a random \cc{shift}. Indeed, \rL{let $Y$ be an independent of
$\Theta$ random variable with   $\mathcal{N}(0,\sigma^2)$ distribution. We} have the equality in distribution
 \BQN
\label{eSH}
 (X, \Theta) \equaldis (\Theta + Y, \Theta).
\EQN
Consequently, \eqref{eqn:BI} follows immediately by the fact that 
the conditional
random variable $\Theta \lvert (\Theta +Y)= x$ is normally distributed for any $x\inr$.\\
The random shifting in this approach \cc{is} related \cc{to} $\Theta$ \cc{which shifts $Y$}.
The random shift model given in \eqref{eSH} has  natural \rL{extensions. For instance, $Y$ can be a $d$-dimensional normally distributed random vector with $\Theta$ being} some $d$-dimensional random vector; 
a more general case is recently discussed in \cite{kume2012calculation}. Another extension  is to consider $Y$ having an elliptical distribution; see Section 4.

In \rL{ruin (or risk)} theory, realistic stochastic models for claim sizes (or risks)  $X_i,i\ge 1$ should  allow for
dependence among them. \rL{Furthermore,} dependent claim sizes need to have a tractable and transparent dependence structure. In several contributions (see e.g., 
\cite{Denetal2006, DLP} and the references therein)
 \ dependent
claim sizes (or risks) are introduced by resorting to \cc{the dependence structure implied} by the Archimedean copula. 
\rL{Recall that an Archimedean copula in $d$-dimension (denoted by $C_\psi$) is defined by
\BQN\label{e1}
 C_\psi(u_1 \ldot u_d)= \psi(  \sum_{i=1}^d \psi^{-1}(u_i)), \quad u_1 \ldot u_d\in [0,1],
 \EQN
where $\psi$ is called the generator of $C_\psi$ required to be positive, strictly decreasing, and \cJ{continuous} with $\psi(0)=1$ and $ \limit{s}\psi(s)=0$, and $\psi^{-1}(x):=\cob{\inf\{t: \psi(t)\le x\}}$;}
see e.g., \cite{ConstH} and the references therein.

A similar idea was used in the context of ruin theory in \cite{Albetal2011} where conditional on the positive random variable $\Theta$
\BQN\label{albr}
\pk{X_1 > x_1\ldot X_n > x_n \lvert \Theta=\theta}= \Bigl(\prod_{i=1}^n \exp(-  x_i)\Bigr)^\theta
\EQN
holds for any positive constants $\theta, x_1 \ldot x_n$. Proposition 1 of the aforementioned paper shows the link of such dependence \rL{structure} (determined by \eqref{albr}) with the Archimedean copula. In fact, instead of dealing with the conditional random model defined in \eqref{albr} we can consider the following equivalent
random \cc{scale} model
\BQN\label{eq:sh2}
(X_1 \ldot X_n) \equaldis  (Y_1/\Theta \ldot Y_n/\Theta),
\EQN
where $Y_i,i\ge 1$ are independent random variables with unit exponential distribution being further independent of the positive random variable $\Theta$.  Clearly, $(X_1 \ldot X_n) \lvert \Theta= \theta$ has joint survival function given by \eqref{albr}.
 The random scale model \eqref{eq:sh2} is interesting  since it leads to certain simplifications; \rL{see} \cite{ConstH}.

In view of the above discussions, some possible approaches for modelling
 dependent claim sizes (or risks) include: \\
\ \  a) copula-based models;\\
 \qquad b) conditional dependence models;\\  
\qquad c) random scale models.

\rL{Of course these are only a few possibilities  which lead to tractable dependence structures with certain appeal to actuarial applications;
 see also 
\cite{valdez2005tail}, 
 \cite{Denetal2006}, \cite{frees2008hierarchical}, \cite{Friz},  \cite{YangIME}, \cite{HashKor13}, \cite{Chavez} and the references therein.}

Finally, we mention that there are several other aspects of actuarial models 
where random shifting and scaling are intrinsically present. For instance, in \cite{Friz} \rL{a new interesting copula model was studied, which can be alternatively introduced} by
a random scale of independent risks; see the discussion in Section 4.

The principal goal of this contribution is to discuss various aspects of random shift and random scale paradigms in actuarial models.
Our analysis leads to new derivations \cc{and insights} \rL{concerning} the calculation  of the Bayesian premium. Furthermore, we show that modeling claim sizes by a   class of Dirichlet random sequences can be done in the \cc{framework} of a \cc{tractable} random scale model.
Further, we point out that random scaling approach is of interest for modeling large \cJ{losses} as in the setup of \cite{Friz}. As a by-product a new class of $L_P$ Dirichlet random vectors is introduced.

 \coE{Organisation of the paper:
    In Section 2 we consider the Bayesian premium \rL{through certain} random shift model. \cc{Our main \cc{finding} is presented in Section 3 which generalizes Theorem 1 in \cite{ConstH}.  Section 4 is dedicated to discussions and extensions.}}

\section{Credibility Premium in Random Shift Models}
For a given $d$-dimensional distribution function $F$ we define a shift family of distribution functions $F(\x;\vk{\theta})= F(\x- \vk{\theta}), \vk{x}, \vk{\theta}\inr^d$. Typically, 
the assumption  on a loss random vector $\X$ is that
 $\X \lvert \vk{\Theta}= \vk{\theta}$ follows a distribution function parametrised by $\vk{\theta}$, say it follows $F(\x;\vk{\theta})$.
 A direct way to formulate this model is via the random shift representation
\BQN\label{rigel}
 (\X, \vk{\Theta})\equaldis ( \vk{Y}+ \vk{\Theta}, \vk{\Theta}),
 \EQN
 where $\vk{Y}$ has distribution function \cc{$F$} and is independent of $\vk{\Theta}$. If $\vk{\Theta}$ \cc{possesses} a \cc{probability density function (pdf)} $h$, then clearly $\X$ also \cc{possesses} a  pdf given by $\E{h(\x- \vk{Y})}$. Consequently, the Bayesian premium (under a $L_2$ loss function) when it exists,  is given by
\BQN
 \E{\vk{\Theta} \lvert \X=\x}&=&  \E{\vk{\Theta} \lvert (\vk{\Theta} + \Y)=\x} \notag\\
 &=& \x -\frac{\E{\vk{Y}h(\vk{x}- \vk{Y})}}{\E{h(\vk{x}- \vk{Y})}}, \label{shift}
 \EQN
where for the derivation of the last equality \eqref{shift} we \cc{assumed} additionally that
$\vk{Y}$ also possesses a \cc{pdf}. Clearly, if $\vk{Y} \equaldis - \vk{Y}$ we have further
\BQN
\E{\vk{\Theta} \lvert \X=\x} = \x + \frac{\E{\vk{Y}h(\vk{x}+ \vk{Y})}}{\E{h(\vk{x}+ \vk{Y})}}.
\EQN

\COM{ be a shift family of $d$-dimensional distribution functions so that
If $Z_1 \ldot Z_d$ are independent $\mathcal{N}(0,1)$ random variables, then the $d$-dimensional random vector $\vk{O}$ with
 \BQN
 \vk{O}\equaldis \Biggl( \frac{Z_1}{(\sum_{i=1}^d\abs{Z_i}^p)^{1/p}}\ldot \frac{Z_d}{(\sum_{i=1}^d \abs{Z_i}^p)^{1/p}}\Biggr), \quad p=2
 \EQN
 is uniformly distributed on unit sphere of $\R^{d}$ with respect to $L_2$ norm.  For $\vk{\Theta}$ a $d$-dimensional random vector with some probability density function $h$ consider the random shift model
 $$ \X= \Y + \vk{\Theta}, \quad \vk{Y}= A R \vk{O},$$
 were $R>0$ is some random variable independent of $\vk{O}$ and $A \inr^{d\times d}$ is a given matrix. When $R^2$ has chi-square distribution with $d$ degrees of freedom, then $R \vk{O} \equaldis (Z_1 \ldot Z_d)$. If we do not specify the distribution function of $R$, then $\vk{Y}$ is an elliptically symmetric random vector. When $R$ possesses a probability density function $q$, then $q$ defines a unique generator $g$ such that $\vk{V}=R \vk{O}$ also possesses a probability density function $f$ given by
\BQN
f(\x) &  =& \frac{1}{c} g\Bigl(\frac{\x^ \top \x}{2}\Bigr), \quad \forall \x\inr^d,
\EQN
where the positive measurable function $g$ is such that
\BQNY
c&=& \frac{(2 \pi)^{d/2} }{\Gamma(d/2)} \int_0^{\IF} u^{d/2-1} g(\cEE{u})\, du \in (0,\IF)
\EQNY
see e.g., \cite{hamada2008capm}, \cite{MR2984355}.\\
The recent paper \cite{kume2012calculation} considers the conditional model that $\vk{X} \lvert \vk{\Theta}= \vk{\theta}$
has the same distribution as $\vk{\Y}+ \vk{\theta}$. 
Clearly, this conditional model is the same as the random shift model $(\Theta + \vk{Y}, \Theta)$, and thus using now the
equivalent representation of the conditional model as a random shift model, the Bayes premium is given by
\BQN \E{ \vk{\Theta} -\X \lvert \X= \x}= \frac{\E{\vk{Y}h(\vk{x}+ \vk{Y})}}{\E{h(\vk{x}+ \vk{Y})}}, \label{shift}
\EQN
provided that the expectation above is finite. Making use of \eqref{shift} we have the following generalisation of Theorem 3.2 in \cite{} for $h$ satisfying
\BQN\label{abs}
h(\x+\vk{z})&= &h(\x)+ \int_0^1 \vk{z}^\top  \nabla h(\x+t\vk{z}) dt, \quad \forall \x,\vk{z} \in \R^d,
\EQN
where $\nabla h$ is the vector function of component-wise partial derivatives, and it is almost surely unique.

We state our first result for $A,R, \vk{V},\vk{Y}, \Theta$. Below the random vector
$\widetilde V= \widetilde R \vk{O}$
where $\widetilde{R}>0$ is independent of $\vk{O}$ has density generator $\widetilde g$ defined by $ \widetilde{g}(x)= \int_x^\IF g(s) \, ds.$
\def\mh#1{\kal{M}_h( #1)}
\def\nh#1{\kal{L}_h( #1)}

\BT Suppose that  $h$ satisfies \eqref{abs}. If further $\E{R^2} < \IF$ and
\BQN
\cEE{\mh{\x}}:=\E{h(\x+\vk{Y})} {\in (0, \IF)}, \quad \E{\abs{\nabla h(\x+A {\vk{\widetilde V}} )}}< \IF,
\EQN
 then the Bayes premium is
\BQN\label{eq:thm:1}
 \E{\vk{\Theta}\lvert \vk{X}=\x}= \x+ c_d A
 \frac{ \E{  \vk{\widetilde V} \nabla h(  \x+ A \vk{\widetilde V} )} }{ \E{ h(\x+ A \vk{\widetilde V})}}, \quad c_d=\frac{ \E{R^2}}{d}.
 \EQN
\ET

 \cite{Garcia92}
}
The random shift model \eqref{rigel} is transparent and offers \cJ{a} clear advantage \rL{in comparison with} the conditional model, if the joint distribution of \rL{$(\vk{\Theta}+ \vk{Y},\vk{\Theta})$ (or $(\vk{\Theta}+ \vk{Y},\vk{Y})$)} can be easily found as illustrated below.

{\bf Example 1}. Suppose that $\X\lvert \vk{\Theta} \sim  \mathcal{N}_d({\vk{\Theta}}, \Sigma)$ \rL{with}
${\vk{\Theta}}\sim \mathcal{N}_d(\vk{\mu}, \Sigma_0)$ (here $\mathcal{N}_d(\vk{\nu}, A)$ stands for the $d$-dimensional normal distribution with mean $\vk{\nu}$ and covariance matrix $A$). \rL{Suppose further that $\Sigma+ \Sigma_0$ is positive definite. 
It follows that $( \vk{X},\vk{\Theta})\equaldis \vk{Z}=(\vk{\Theta}+ \vk{Y},\vk{\Theta})$ with $\vk{Y}
\sim \mathcal{N}_d(\vk{0}, \Sigma)$ independent of $\vk{\Theta}$. Therefore, in the light of \cite{Denetal2006} 
 the fact that $\vk{Z}$ is normally distributed in  $\R^{2d}$ implies that $\vk{Y} \lvert (\vk{\Theta}+\vk{Y})= \vk{x}$
is normally distributed with mean
$$ \vk{\bar \mu}=\E{\vk{Y} \lvert (\vk{\Theta}+\vk{Y}) =\x}=(\x- \vk{\mu})( \Sigma+ \Sigma_0)^{-1} \Sigma.$$
Consequently 
\BQN\label{licht0}
\E{\vk{\Theta} \lvert \X=\x}=\x-\E{\vk{Y} \lvert (\vk{\Theta}+\vk{Y}) =\x}= \x+ ( \vk{\mu}-\x)( \Sigma+ \Sigma_0)^{-1} \Sigma.
\EQN
Particularly, if $\Sigma$ is positive definite
\BQN\label{licht}
\E{\vk{\Theta} \lvert \X=\x}= \x+ ( \vk{\mu}-\x) ( \Sigma_0\Sigma^{-1} + I_d)^{-1},
\EQN
where  $I_d$ denotes the $d\times d$ identity matrix.}

Clearly, \cc{(1) is immediately established by the above} for the special case that $d=1$ and $\Sigma=\sigma^2,\Sigma_0=\tau^2 $ .\\
It is worth pointing out that \eqref{licht} was derived by \cite{kume2012calculation} when $\Sigma_0$ is non-singular using an indirect (in that case complicated) approach; whereas Example 1 gives a short direct proof for the formula of the Bayesian premium in the random shift Gaussian model, where we can further allow $\Sigma_0$ to be singular.

\section{Dirichlet Claim Sizes $\&$  Random Scaling}
\cc{A fundamental question when constructing models for claim sizes $X_i,i\ge 1$ is how to introduce tractable dependence structures.}
As mentioned in the Introduction, one common approach in the actuarial literature is to assume that the survival copula of $X_i, i\le n$ is a $n$-dimensional Archimedean copula; see e.g., 
\cite{wu2007simulating}, \cite{Albetal2011} and the references therein. In view of the link between  Archimedean copula and Dirichlet distribution explained in \cite{McNNev2009}, we choose the direct approach for modeling claim sizes by a Dirichlet random sequence \rL{as in \cite{ConstH}}.

 \rL{With motivation from the definition of $L_1$ Dirichlet random vectors, we  introduce next} $d$-dimensional $L_p$ Dirichlet random vectors. Let $Gamma(a,\lambda)$ denote the Gamma distribution with positive parameters $a,\lambda$. It is known that the  pdf  of it is $\lambda^a x^{a-1} \exp(- \lambda x)/\Gamma(a)$, where $\Gamma(\cdot)$ stands for the Euler Gamma function.
Fix some positive constants $\alpha_i,i\ge 1$, and $p$. In the rest of the paper, without special indication,  let $Y_i, i\ge 1$ denote a sequence of positive independent random variables  defined on some probability space  $(\Omega, \mathcal{A}, {\bf P})$ such that, for any $i\ge1,$ $Y_i^p$ \rL{has $Gamma(\alpha_i,1/p)$ distribution}  with parameters  $\alpha_i$ and $p$. It follows easily that the \cJ{pdf} of $Y_i$ is given by
$$
f_{i}(x)=\frac{p^{1-\alpha_i}}{\Gamma(\alpha_i)} x^{p \alpha_i -1} \exp\left(- \frac{x^p}{p}\right),\ \ \  x>0.
$$
We say that $(X_1 \ldot X_d)$ is a $d$-dimensional $L_p$ Dirichlet random vector, if
 the  stochastic representation
\BQN \label{OO}
(X_1 \ldot  X_d)&\equaldis &\left(
 R \frac{Y_1}{(\sum_{1 \le i \le d} Y_i^p )^{1/p}} \ldot  R\frac{Y_d}{(\sum_{1 \le i \le d} Y_i^p)^{1/p}}\right)=:R \vk{O}
\EQN
holds with some positive random variable $R$  defined on  $(\Omega, \mathcal{A}, {\bf P})$ which is independent of the random vector $\vk{O}$.
The reason for the name of  $L_p$ Dirichlet random vector (and distribution) is that the angular component $\vk{O}$ lives on the unit $L_p$-sphere of $\R^d$, i.e.,
$$ \sum_{i=1}^d O_i^p= 1.$$
When $p=1$, $\vk{O}$ has the Dirichlet distribution on the unit simplex; see \cite{{McNNev2009}}.
\COM{ Clearly, $X_i,X_j$ for any two different indices $i$ and $j$ are in general dependent random variables. Independence of components of $\vk{X}=(X_1 \ldot  X_d)$ is also possible when  $R^p$ has $Gamma(\sum_{i=1}^d\alpha_i, 1/p)$ distribution. For instance, the random vector $\vk{Y}=(Y_1 \ldot Y_d)$ is  a Dirichlet random vector with independent components and stochastic representation $\vk{Y} \equaldis \tilde R \vk{O}$, where  $\tilde R \ge 0$ being independent of $\vk{O}$ is such that $\tilde R^p \equaldis \sum_{i=1}^d Y_i^p$, i.e., $\tilde R^p$ has $Gamma(\sum_{i=1}^d\alpha_i, 1/p)$ distribution.}

\COM{Since in numerous models an infinite sequence $X_i,i\ge 1$ of random variables is considered to have a common underlying distribution function $F$, we shall assume next that all $\alpha_i'$s are the same equal to some $\alpha>0$, and therefore the $i$th component $O_i$ of the random vector $\vk{O}$ is such that $O_i^p$ has Beta distribution with parameters $\alpha$ and $d \alpha$.\\}

\rL{The main result of this section} displayed in the next theorem shows that the model with Dirichlet claim sizes can be explained by a random scale model.

\BT \label{prop:1} Let $X_i, i\ge 1$ be positive random variables. 
If, for any $d\ge 2$, the random vector $(X_1 \ldot X_d)$ has
a $d$-dimensional $L_p$ Dirichlet distribution with representation $(X_1 \ldot X_d)\equaldis  R_d \vk{O}_d$, then
\BQN\label{licht2}
\{X_i, i\ge  1\} \equaldis \{S Y_i,i\ge1\},
\EQN
with $S$ a non-negative random variable defined on  $(\Omega, \mathcal{A}, {\bf P})$, independent of $Y_i, i\ge 1$. 
\ET

{\bf Proof}: By definition, it is sufficient to show that, for any $d\ge1$
\BQN\label{XY}
(X_1 \ldot X_d) \equaldis S(Y_1 \ldot Y_d)
\EQN
for the non-negative random variable $S$ required.
Since for any $n\ge d$ the random vector $(X_1 \ldot X_n)$ has a $L_p$ Dirichlet distribution, then we have the stochastic representation
\BQN \label{id}
(X_1 \ldot X_d) \equaldis \frac{R_n}{a_n} \frac{1}{(\sum_{i=1}^nY_i^p)^{1/p}/a_n}(Y_1 \ldot Y_d),
\EQN
with $ a_n= (\sum_{i=1}^n \alpha_i)^{1/p}$. Therefore, we have \cJ{the convergence in distribution (denoted here as $\todis$)}
\BQNY \label{idd}
\frac{R_n}{a_n} \frac{1}{(\sum_{i=1}^nY_i^p)^{1/p}/a_n}(Y_1 \ldot Y_d)  \todis(X_1 \ldot X_d)
\EQNY
as $n\to\IF$.
Clearly, by the strong law of large numbers, as $n \to \infty$ we have the almost sure
convergence $(\sum_{i=1}^n Y_i^p)^{1/p}/a_n \to 1 $ which entails
\BQNY
\frac{R_n}{a_n} (Y_1 \ldot Y_d)  \todis(X_1 \ldot X_d)
\EQNY
as $n\to \IF$, meaning that 
$$\ln \left(\frac{R_n}{a_n}\right) + (\ln (Y_1)\ldot \ln (Y_d)) \todis (\ln (X_1)\ldot \ln(X_d)), \quad n\to \IF.$$ In the light of Theorem 3.9.4 in \cite{Durrett2010}, by the independence of $R_n$ and $(Y_1 \ldot Y_d)$ we conclude that
$$ \frac{R_n}{a_n} \todis S, \quad n\to \IF, $$
with $S$ some non-negative random variable defined on  $(\Omega, \mathcal{A}, {\bf P})$  such that
$$  (\ln (Y_1)+\ln \left(S\right)\ldot \ln (Y_d)+\ln \left(S\right)) \equaldis (\ln (X_1)\ldot \ln(X_d))$$
implying \eqref{XY}, and thus the claim follows. \QED

\rL{The following corollary is a generalization of Theorem 1 in \cite{ConstH}.}

\BK If the claim sizes $X_i, i\ge 1$ are identically distributed, then  under the assumptions and notation of \netheo{prop:1} \eqref{licht2}
holds with $Y_i,i\ge 1$ a sequence of independent random variables with common \cJ{pdf} $f(x)= p^{1-\alpha}/\Gamma(\alpha) x^{p \alpha -1}
 \exp(- x^p/p), x>0$, for some $\alpha >0$.
\EK
In view of the well-known Beta-Gamma algebra \rL{(see e.g., \cite{YorM})} if $\alpha \in (0,1)$, then $Y_i$ in Corollary 2 can be re-written as
$$ Y_i \equaldis (T_i E_i)^{1/p}, \quad i\ge 1,$$
with $T_i$ \rL{a} Beta distribution with parameters $\alpha, 1- \alpha$ and \rL{$E_i$ being exponential distributed with mean $p$}. Further, $T_i,E_i,i\ge 1$
are mutually independent. Consequently
$$ (X_1 \ldot X_n) \equaldis (S (T_1 E_1)^{1/p}\ldot S (T_n E_n)^{1/p}), \quad n \ge 1.$$
\cc{Note that $(S E_1 \ldot S E_d), d>1$ is a $d$-dimensional $L_1$ Dirichlet random  vector.}


\section{Discussions \& Extensions }
The conditional credibility model that we considered in Section 2 is simple since we used a single distribution function $F$ to define a shift family of distributions, i.e., $F(\x, \vk{\theta})= F(\x - \vk{\theta})$. Of course, we can consider \cJ{a more general case that $F=F_{\vk{\theta}}$ is a family of $d$-dimensional distributions} and assume that $\X \lvert \vk{\Theta}= \vk{\theta}$ has distribution function $F_{\vk{\theta}}(\x- \vk{\theta})$.
Hence the random shift model is $(\X, \vk{\Theta}) \equaldis (\vk{\Theta} + \Y, \vk{\Theta})$, where $\Y \lvert \vk{\Theta}=  \vk{\theta}$
 has distribution function $F_{\vk{\theta}}$. It is clear that the random shift model is again specified via a conditional distribution, so there is no essential simplification by re-writing the conditional model apart from the case that the joint distribution of $(\vk{Y}, \vk{\Theta})$ is known.\\
We consider briefly a tractable instance  that $(\Y,\vk{\Theta})$ has an elliptical   distribution in $\R^{2d}$, i.e.,
\BQN\label{Ellip}
(\Y,\vk{\Theta}) \equaldis \Biggl(R \frac{Z_1}{\sqrt{\sum_{i=1}^{2d} Z_i^2}} \ldot R \frac{Z_{2d}}{\sqrt{\sum_{i=1}^{2d} Z_i^2}}\Biggr)  C + \vk{\nu}=: R \vk{U}C+\vk{\nu}, \quad \vk{\nu}\inr^{2d},
\EQN
 with $Z_i,i\le \cb{2d}$ independent $\mathcal{N}(0,1)$ distributed random variables being further independent of $R>0$, and $C$ a square matrix in $\R^{2d \times 2d}$. For more details and actuarial applications of elliptically symmetric multivariate distributions see \cite{Denetal2006}.\\
Let  $I=\{1 \ldot d\}$ and $J=\{d+1 \ldot 2d\}$. For any $2d\times2d$ matrix $A$, denote $A_{I,J}$ as the sub-matrix of $A$ obtained by selecting the elements with row indices in $I$ and column indices in $J$. Similarly, for any row vector $\vk{\nu}\in\R^{2d}$, define $\vk{\nu}_I$ and  $\vk{\nu}_J$ to be the sub-vectors of $\vk{\nu}$. Further, denote by $A^{\top}$ the transpose of matrix $A$.\\
 \rL{By the stochastic representation \eqref{Ellip} we obtain that   
$$(\vk{\Theta}+\Y,\vk{\Theta}) \equaldis  R \vk{U}C^*+ \vk{\nu}^*,$$
 where
 $$
 C^*=\left(
          \begin{array}{cc}
            C_{I,I}+C_{I,J}& C_{I,J} \\
            C_{J,I}+C_{J,J} & C_{J,J} \\
          \end{array}
        \right),\ \ \ \ \ \ \vk{\nu}^*=(\vk{\nu}_I+ \vk{\nu}_J,  \vk{\nu}_J).
 $$ 
  Set $B= (C^*)^\top C^* $ and assume that $B$ is non-singular.
  As in the Gaussian case, for the more general class of elliptically symmetric distributions, the conditional random vector $\vk{\Theta} \lvert ( \vk{\Theta} + \vk{Y}) =\x$ is again elliptically symmetric with stochastic representation (suppose for simplicity $\vk{\nu}_I=\vk{0}, \vk{\nu}_J=:\vk{\mu}$)
  $$
   \vk{\Theta} \lvert ( \vk{\Theta} + \vk{Y}) =\x \equaldis \vk{\mu}+\cbb{(\x - \vk{\mu})}B_{I,I}^{-1}B_{I,J}   + R_{\vk{x}} \vk{U}D,
   $$
   where $D$ is a square matrix such that $D^\top D =B_{J,J}- B_{J,I} B_{I,I}^{-1} B_{I,J}$, and the random variable $R_{\vk{x}}>0$ is 
   independent of $\vk{U}$; see e.g., \cite{Cam1981}. Consequently, since $\vk{U}$ has components being symmetric about 0,
   we obtain the Bayesian premium  formula
   \BQN
   \E{\vk{\Theta}\lvert \X=\x}= \vk{\mu}+ \cbb{(\x - \vk{\mu})}B_{I,I}^{-1}B_{I,J},
   \EQN
    provided that $\E{R_{\x}}< \IF$.
In the special case that $C_{I,J}$ and $C_{J,I}$ have  all entries equal to 0, and further
 $$C_{I,I}^\top C_{I,I}=\Sigma \quad C_{J,J}^\top C_{J,J}=\Sigma_0,$$
we conclude that  \eqref{licht0} holds.} 

The random vector $\vk{O}$ defined in \eqref{OO} has components $O_i,i\le d$ such that $O_i^p$ has beta distribution with parameters $\alpha_i, \sum_{j\le d, j\not =i} \alpha_j$; see e.g., \cite{ConstH}.  In the special case that $\alpha_i=1/p$ for any $i\le d$, properties of $\vk{O}$ and $\X= R \vk{O}$ with $R>0$ independent of $\vk{O}$
are studied in \cite{szab}. Our result in Corollary 2 agrees with the finding of Theorem 4.4  in the aforementioned paper.
Note that, for the case $p=2$ the corresponding result of \netheo{prop:1} for spherically  symmetric random sequences is well-known, see e.g., \citet{Sch1938} or \citet{Bry1995}.

 \cob{Weighted} $L_p$ Dirichlet random vectors are \cc{naturally} introduced by using \coE{indicator} random variables. \cc{Specifically},
\rL{let $R>0$ and $\vk{O}$ be given as in \eqref{OO}. Further, let $I_i,i\le d$ be independent Bernoulli random variables defined on $(\Omega, \mathcal{A}, {\bf P})$ with $\pk{I_i=1}=\coE{q_i}=1-\pk{I_i=-1},q_i\in (0,1], i\le d$, which is further independent of the random vector $(R,\vk{O})$. }\coB{The random vector $ \X$ with stochastic representation}
\BQN \label{eq:symYi}
\X &\equaldis &(R I_1 O_1 \ldot R  I_d O_d)
 \EQN
 is  referred to as a weighted $L_p$ Dirichlet random vector with indicators $I_i,i\le d$ and parameters $\alpha_1 \ldot \alpha_d$. \\
If $I_i$'s are iid with $\E{I_1}=0$ and $\alpha_1 = \ldots = \alpha_d=2=p,$ then
$$
( I_1 O_1 \ldot   I_d O_d) \equaldis \Biggl(\frac{Z_1}{\sqrt{\sum_{i=1}^d Z_i^2}} \ldot \frac{Z_{d}}{\sqrt{\sum_{i=1}^{d} Z_i^2}}\Biggr).
$$
Therefore, if  $R^2$ is chi-square distributed with $d$ degrees of freedom then $\X$ has a centered Gaussian distribution with $N(0,1)$ independent components. The introduction of the  weighted Dirichlet random vectors is important since it includes  the normal distribution as a special case. In addition, weighted Dirichlet random vectors are suitable for modeling claim sizes in certain ruin models with double-sided jumps; see Example 4 in \cite{ConstH}.


As in the case of $L_p$ Dirichlet random sequences,  in the weighted case the dependence structure can be given through a random scale model as well. More precisely, if the random sequence $X_i,i\coC{\ge} 1,$ is such that, for any fixed \cJ{$d\ge2$}, $(X_1 \ldot X_d)$ is a weighted $L_p$ Dirichlet random vector with indicators $I_i,i\le d$ and parameters $\alpha_1 \ldot \alpha_d$, then 
\BQN\label{infWD}
\{ X_i, i\ge1\} &\coE{\equaldis} & \{S I_i  Y_i, i\ge1\},
\EQN
with $\cEE{S}$ some non-negative random  variable defined on $(\Omega, \mathcal{A}, {\bf P})$ which is independent of 
$I_i,Y_i,i\ge 1$.

With motivation \cc{from} credibility theory, we propose to consider 
a new class of multivariate distributions called $L_P$ Dirichlet distributions, which is \cc{naturally introduced} by letting the parameter $p$ in our definition above to be random \cc{(a common feature of credibility models where parameters are random elements)}.\\
Specifically, let $P$ be a positive random variable, and let $\alpha_i, i\ge1$ be positive constants.
Further, let $Y_i,i\ge1$ be independent random variables, which \cc{are} further independent of $P$ such that \cc{$Y_i^P$}
has $Gamma(\alpha_i,1)$ distribution. We say that $(X_1 \ldot X_d)$
is a $d$-dimensional $L_P$ Dirichlet random \cJ{vector} if 
\BQN \label{OOP}
(X_1 \ldot  X_d)&\equaldis &\left(
 R \frac{Y_1}{(\sum_{1 \le i \le d} Y_i^P )^{1/P}} \ldot  R\frac{Y_d}{(\sum_{1 \le i \le d} Y_i^P)^{1/P}}\right)=:R \vk{O}^{(P)}
\EQN
holds for some positive random variable $R$ independent of $P$ and $(Y_1 \ldot Y_d)$.
Here, for any $\vk{O}^{(P)}=(O_1\ldot O_d)$
$$
\sum_{i=1}^d O_i^P=\cJ{1.} 
$$
This multivariate distribution can be used in the context of credibility models, models for large losses, models for risk aggregation,
 or \rL{models for} claim sizes.
If we assume that $X_i,i\ge 1$ is a sequence of claim sizes such that, for any $n\ge2$, $(X_1 \ldot X_d)$ has a $d$-dimensional $L_P$ Dirichlet distribution, then an extension of Theorem 1 for this case is possible.
\rL{More precisely 
 choosing now $a_n=(\sum_{i=1}^n\alpha_i)^{1/P}$ we conclude} that
$$ \{X_i,i\ge1\}\equaldis \{S Y_i,i\ge1\},$$
with some non-negative random variable $S$ independent of \rL{$Y_i, i\ge1$.}  

To this end, we consider a new copula class introduced in \cite{Friz} which is referred to as MGB2 copula. Let $\Theta$ be a positive random variable having an inverse Gamma distribution with shape parameter $q>0$ and a unit scale, i.e., $1/\Theta$ has $Gamma(q,1)$ distribution. In view of the aforementioned paper $(X_1 \ldot X_n)$ has a MGB2 distribution (or MGB2 copula) if $X_i$'s are positive random variables and $X_1 \lvert \Theta = \theta \ldot X_n \lvert \Theta = \theta$ are independent with \cJ{pdf}
$f_{X_i\lvert \Theta}, i\le n$ given by
    $$f_{X_i|\Theta}(x_i|\theta)=\frac{a_i}{\Gamma(p_i)x_i\theta^{p_i}}\left(\frac{x_i}{b_i}\right)^{a_ip_i}e^{-\theta^{-1}(x_i/b_i)^{a_i}}, \qquad x_i>0, \theta>0.$$
 Here the parameters $ a_i,b_i,p_i,i\le n$ are all positive constants.
Instead of using the conditional argument, we can directly define $(X_1 \ldot X_n)$ via a random scale model as follows
\BQN \label{fri}
(X_1 \ldot X_n) \equaldis (\Theta^{1/a_1} W_1 \ldot \Theta^{1/a_n} W_n),
\EQN
with $W_1 \ldot W_n$ being independent positive random variables such that, for any fixed $i\le n$, $W_i^{a_i}$ has $Gamma(p_i,b_i^{-a_i} )$ distribution.  One advantage of the random scale model \eqref{fri}
is that, for modeling purposes, \rL{it can be re-written} as a random shift model
\BQN \label{fri2}
(\ln X_1 \ldot \ln X_n) \equaldis \Bigl(\frac{1}{a_1} \ln \Theta +\ln W_1 \ldot \frac{1}{a_n} \ln \Theta + \ln W_n \Bigr).
\EQN
\rL{Another} advantage of the random scale model \eqref{fri} becomes clearer if of interest is the joint tail asymptotic behaviour of $(X_1, X_2)$, \rL{as discussed in \cite{Friz}. As illustrated below, the regular variation of survival function of $\Theta$ is enough for the joint tail asymptotic behaviour of $(X_1, X_2)$; distributional assumptions on $\Theta$ are not really necessary.}

{\bf Example 2}.  Let $(W_1,W_2)$ be defined as above with the parameters therein and further assume that $a_1=a_2=a>0$.
  \rL{Define  $(X_1,X_2)$ through \eqref{fri} with $\Theta$ an independent of $(W_1,W_2)$ random variable having a regularly varying tail behavior at infinity with index $q>0$, i.e.,
  $$\lim_{x\to \IF} \frac{\pk{\Theta> tx}}{\pk{\Theta > x}}= t^{-q}, \ \ \forall t> 0.$$
  For modeling joint behaviour of large \cJ{losses} of interest is the calculation of the following limit
$$ \limit{t} \frac{ \pk{X_1> c_1 t, X_2 > c_2t}}{ \pk{X_1> t}}$$ 
for $c_1,c_2$ positive constants, see e.g., \cite{Cden} and \cite{Denetal2006}.} In our case we have
\BQNY
\frac{\pk{X_1> c_1 t, X_2 > c_2t}}{\pk{X_1> t}} &=& \frac{\pk{\Theta^{1/a} W_1/c_1 >  t, \Theta^{1/a} W_2/c_2 > t}}{\pk{\Theta^{1/a} W_1> t}}\\
&=&  \frac{\pb{\Theta \min \Bigl( (W_1/c_1)^{a},(W_2/c_2)^{a} \Bigr)  > t^{a}}}{\pk{\Theta W_1^{a}> t^{a}}}\\
&\to&  \frac{\eg{ \Bigl(\min( W_1/c_1,W_2/c_2) \Bigr)^{aq}}}{\E{ W_1^{aq}}}=I(c_1,c_2)>0
\EQNY
as $t\to \IF$ where in the last step we applied Breiman's lemma; see e.g., \cite{MR2271424}. Since the asymptotic dependence function $I(c_1,c_2)$ is positive,
an appropriate extreme value model for the joint survival function of $X_1$ and $X_2$ is the one that allows for \FRE  marginals and asymptotic dependence.

\section{Conclusion}
This contribution shows that in various insurance applications besides common conditional stochastic models, equivalent
random shift or random \cc{scale} models can be analysed \cc{and explored}.
As explained in the context of credibility models, simple random shift models lead  to direct derivations
for \cc{the calculation} of the Bayesian premium. In particular, Example 1 shows that for Gaussian models, the covariance matrix of the prior distribution can be singular without changing the outcome.\\
Our main result concerning the random scale property of $L_p$ Dirichlet random sequences is not only of theoretical importance but also of practical importance, since in certain models claim sizes can be reduced to random scale of independent random sequences with known marginal distributions. \\
Example 2 demonstrates the usefulness of random scale models for analysing joint survival functions for large thresholds.
\cc{As a by-product in Section 4 we suggest a new dependence structure of interest for dependent risks.}\\

\textbf{Acknowledgments.} \cb{Partial support from the Swiss National Science Foundation Project 200021-140633/1  and RARE -318984
 (an FP7 Marie Curie IRSES Fellowship) is kindly acknowledged.}

\bibliographystyle{plainnat}
\bibliography{Arch3}

\end{document}